\documentclass{article}

\def\abstract#1{\vskip 7mm 
        \begin{center}{\large Abstract}\par \smallskip
                \begin{minipage}[c]{12cm}
                        \small #1
                \end{minipage}
        \end{center}
}
\def\title#1{\begin{center}{\Large\bf #1}\end{center}}
\def\author#1{\vskip 5mm \begin{center}{#1}\end{center}}
\def\address#1{\begin{center}{\it #1}\end{center}}
\makeatletter
\@ifundefined{lesssim}{}{}
\@ifundefined{gtrsim}{}{}
\def\vereq#1#2{\lower3pt\vbox{\baselineskip1.5pt \lineskip1.5pt
\ialign{$\m@th#1\hfill##\hfil$\crcr#2\crcr\sim\crcr}}}
\makeatother

\begin{document}

\title{%
  Inclusion of the first-order vector- and tensor-modes in the
  second-order gauge-invariant cosmological perturbation theory 
}
\author{%
  Kouji Nakamura\footnote{E-mail:kouchan@th.nao.ac.jp}
}
\address{%
  Department of Astronomical Science,
  the Graduate University for Advanced Studies,\\
  Mitaka 181-8588, Japan
}

\abstract{
  Gauge-invariant treatments of the second-order cosmological
  perturbation in a four dimensional homogeneous isotropic
  universe are formulated without any gauge fixing.
  We have derived the Einstein equations in the case of the
  single perfect fluid without ignoring any modes.
  These equations imply that any types of mode-coupling arise
  due to the second-order effects of the Einstein equations.
}


The second-order general relativistic cosmological perturbation
theory has very wide physical motivation.
In particular, the first order approximation of our universe
from a homogeneous isotropic one is revealed by the recent
observations of Cosmic Microwave Background (CMB) by Wilkinson
Microwave Anisotropy Probe\cite{WMAP}, which suggests that the 
fluctuations of our universe are adiabatic and Gaussian at least
in the first order approximation.
One of the next theoretical researches is to clarify the
accuracy of this result through the non-Gaussianity, or
non-adiabaticity, and so on.
To carry out this, it is necessary to discuss the second-order
cosmological perturbations.


However, general relativistic perturbation theory requires
delicate treatments of ``gauges'' and this situation becomes
clearer by the general arguments of perturbation theories.
Therefore, it is worthwhile to formulate the higher-order
gauge-invariant perturbation theory from general point of view. 
According to this motivation, we proposed the general framework
of the second-order gauge-invariant perturbation theory on a
generic background spacetime\cite{KNs-general}.
This general framework was applied to cosmological perturbation
theory\cite{KNs-cosmological} and all components of the
second-order perturbation of the Einstein equation were derived
in gauge invariant manner.
The derived second-order Einstein equations are quite similar to
the equations for the first-order one but there are source terms
which consist of the quadratic terms of the linear-order
perturbations.


In this article, we show the extension of the formulation in
Refs.~\cite{KNs-cosmological} to include the first-order vector-
and tensor-modes in the source terms of the second-order
Einstein equation, which were ignored in
Refs.~\cite{KNs-cosmological}.


As emphasized in Refs.\cite{KNs-general,KNs-cosmological}, in
any perturbation theory, we always treat two spacetime manifolds. 
One is a physical spacetime ${\cal M}_{\lambda}$ and the other
is the background spacetime ${\cal M}_{0}$.
In this article, the background spacetime ${\cal M}_{0}$ is the
Friedmann-Robertson-Walker universe filled with a perfect fluid
whose metric is given by
\begin{eqnarray}
  g_{ab} = a^{2}(\eta)\left(
    - (d\eta)_{a}(d\eta)_{b}
    + \gamma_{ij}(dx^{i})_{a}(dx^{j})_{b}
  \right),
  \label{eq:background-metric}
\end{eqnarray}
where $\gamma_{ij}$ is the metric on maximally symmetric three
space.
The physical variable $Q$ on the physical spacetime is pulled
back to ${}_{{\cal X}}\!Q$ on the background spacetime by 
an appropriate gauge choice ${\cal X}$ which is an
point-identification map from ${\cal M}_{0}$ to 
${\cal M}_{\lambda}$.
The gauge transformation rules for the pulled-back variable
${}_{{\cal X}}\!Q$, which is expanded as 
${}_{{\cal X}}\!Q_{\lambda}$ $=$ $Q_{0}$
$+$ $\lambda {}^{(1)}_{{\cal X}}\!Q$
$+$ $\frac{1}{2} \lambda^{2} {}^{(2)}_{{\cal X}}\!Q$, are given by 
\begin{eqnarray}
  \label{eq:Bruni-47-one}
  {}^{(1)}_{\;{\cal Y}}\!Q - {}^{(1)}_{\;{\cal X}}\!Q = 
  {\pounds}_{\xi_{(1)}}Q_{0}, \quad
  {}^{(2)}_{\;\cal Y}\!Q - {}^{(2)}_{\;\cal X}\!Q = 
  2 {\pounds}_{\xi_{(1)}} {}^{(1)}_{\;\cal X}\!Q 
  +\left\{{\pounds}_{\xi_{(2)}}+{\pounds}_{\xi_{(1)}}^{2}\right\} Q_{0},
\end{eqnarray}
where ${\cal X}$ and ${\cal Y}$ represent two different gauge choices,
$\xi_{(1)}^{a}$ and $\xi_{(2)}^{a}$ are generators of the first- and
the second-order gauge transformations, respectively.
The metric $\bar{g}_{ab}$ on the physical spacetime 
${\cal M}_{\lambda}$ is also expanded as
$\bar{g}_{ab}$ $=$ $g_{ab}$ $+$ $\lambda h_{ab}$ $+$ 
$\frac{\lambda^{2}}{2} l_{ab}$ under a
gauge choice.
Inspecting gauge transformation rules
(\ref{eq:Bruni-47-one}), the first-order
metric perturbation $h_{ab}$ is decomposed as 
$h_{ab}$ $=:$ ${\cal H}_{ab}$ + ${\pounds}_{X}g_{ab}$, where
${\cal H}_{ab}$ and $X_{a}$ are transformed as 
${}_{\;{\cal Y}}\!{\cal H}_{ab}$ $-$ 
${}_{\;{\cal X}}\!{\cal H}_{ab}=0$, and 
${}_{\;{\cal Y}}\!X_{a}$ $-$ ${}_{\;{\cal X}}\!X_{a}$ $=$
$\xi_{(1)a}$ under the gauge transformation
(\ref{eq:Bruni-47-one}), respectively\cite{KNs-cosmological}.
The gauge invariant part ${\cal H}_{ab}$ of $h_{ab}$ is given in
the form
\begin{eqnarray}
  \label{eq:first-order-gauge-inv-metrc-pert-components}
  {\cal H}_{ab}
  &=& 
  - 2 a^{2} \stackrel{(1)}{\Phi} (d\eta)_{a}(d\eta)_{b}
  + 2 a^{2} \stackrel{(1)}{\nu_{i}} (d\eta)_{(a}(dx^{i})_{b)}
  + a^{2} 
  \left( - 2 \stackrel{(1)}{\Psi} \gamma_{ij} 
    + \stackrel{(1)}{{\chi}_{ij}} \right)
  (dx^{i})_{a}(dx^{j})_{b},
\end{eqnarray}
where
$D^{i}\stackrel{(1)}{\nu_{i}}$ $=$ $\stackrel{(1)}{\chi_{[ij]}}$
$=$ $\stackrel{(1)}{\chi^{i}_{\;i}}$ $=$
$D^{i}\stackrel{(1)}{\chi_{ij}}$ $=$ $0$ and
$D^{i}:=\gamma^{ij}D_{j}$ is the covariant derivative associate
with the metric $\gamma_{ij}$.
In the cosmological perturbations\cite{Bardeen-1980},
$\{\stackrel{(1)}{\Phi},\stackrel{(1)}{\Psi}\}$,
$\stackrel{(1)}{\nu_{i}}$, and $\stackrel{(1)}{\chi_{ij}}$ are
called the scalar-, vector-, and tensor-modes, respectively.
We have to note that we used the existence of the Green
functions $\Delta^{-1}=:(D^{i}D_{i})^{-1}$, $(\Delta+2K)^{-1}$,
and $(\Delta+3K)^{-1}$ to accomplish the above decomposition of
$h_{ab}$.


As shown in Ref.\cite{KNs-general}, through the above variables
$X_{a}$ and $h_{ab}$, the second order metric perturbation
$l_{ab}$ is decomposed as $l_{ab}$ $=:$ ${\cal L}_{ab}$ $+$ 
$2 {\pounds}_{X}h_{ab}$ $+$ 
$\left( {\pounds}_{Y} - {\pounds}_{X}^{2} \right) g_{ab}$
The variables ${\cal L}_{ab}$ and $Y^{a}$ are the gauge
invariant and variant parts of $l_{ab}$, respectively.
The vector field $Y_{a}$ is transformed as 
${}_{\;{\cal Y}}\!Y_{a}$ $-$ ${}_{\;{\cal X}}\!Y_{a}$ $=$
$\xi_{(2)}^{a}$ $+$ $[\xi_{(1)},X]^{a}$ under the gauge
transformations (\ref{eq:Bruni-47-one}).
The components of ${\cal L}_{ab}$ are given by 
\begin{eqnarray}
  \label{eq:second-order-gauge-inv-metrc-pert-components}
  {\cal L}_{ab}
  &=& 
  - 2 a^{2} \stackrel{(2)}{\Phi} (d\eta)_{a}(d\eta)_{b}
  + 2 a^{2} \stackrel{(2)}{\nu_{i}} (d\eta)_{(a}(dx^{i})_{b)}
  + a^{2} 
  \left( - 2 \stackrel{(2)}{\Psi} \gamma_{ij} 
    + \stackrel{(2)}{{\chi}_{ij}} \right)
  (dx^{i})_{a}(dx^{j})_{b},
\end{eqnarray}
where
$D^{i}\stackrel{(2)}{\nu_{i}}$ $=$ $\stackrel{(2)}{\chi_{[ij]}}$
$=$ $\stackrel{(2)}{\chi^{i}_{\;\;i}}$ $=$
$D^{i}\stackrel{(2)}{\chi_{ij}}$ $=$ $0$. 
As shown in Ref.\cite{KNs-general}, by using the above
variables $X_{a}$ and $Y_{a}$, we can find the gauge invariant
variables for the perturbations of an arbitrary field as
\begin{eqnarray}
  \label{eq:matter-gauge-inv-def-1.0}
  {}^{(1)}\!{\cal Q} := {}^{(1)}\!Q - {\pounds}_{X}Q_{0},
  , \quad 
  {}^{(2)}\!{\cal Q} := {}^{(2)}\!Q - 2 {\pounds}_{X} {}^{(1)}Q 
  - \left\{ {\pounds}_{Y} - {\pounds}_{X}^{2} \right\} Q_{0}.
\end{eqnarray}


As the matter contents, in this article, we consider a perfect
fluid whose energy-momentum tensor is given by
$\bar{T}_{a}^{\;\;b}$ $=$ 
$\left(\bar{\epsilon}+\bar{p}\right)\bar{u}_{a}\bar{u}^{b}$ $+$
$\bar{p}\delta_{a}^{\;\;b}$. 
We expand these fluid components $\bar{\epsilon}$, $\bar{p}$,
and $\bar{u}_{a}$ as
\begin{eqnarray}
  \bar{\epsilon}
  =
  \epsilon
  + \lambda \stackrel{(1)}{\epsilon}
  + \frac{1}{2} \lambda^{2} \stackrel{(2)}{\epsilon}
  , 
  \quad
  \bar{p}
  =
  p
  + \lambda \stackrel{(1)}{p}
  + \frac{1}{2} \lambda^{2} \stackrel{(2)}{p} 
  ,
  \quad
  \bar{u}_{a}
  =
  u_{a}
  + \lambda \stackrel{(1)}{u}_{a}
  + \frac{1}{2} \lambda^{2} \stackrel{(2)}{u}_{a}p 
  .
  \label{eq:Taylor-expansion-of-four-velocity}
\end{eqnarray}
Following the definitions (\ref{eq:matter-gauge-inv-def-1.0}), we
easily obtain the corresponding gauge invariant variables for these
perturbations of the fluid components:
\begin{eqnarray}
  \stackrel{(1)}{{\cal E}} 
  &:=& \stackrel{(1)}{\epsilon} - {\pounds}_{X}\epsilon, \quad
  \stackrel{(1)}{{\cal P}}
  := \stackrel{(1)}{p} - {\pounds}_{X}p, \quad
  \stackrel{(1)}{{\cal U}_{a}}
  := \stackrel{(1)}{(u_{a})} - {\pounds}_{X}u_{a}, \nonumber
 \quad
  \stackrel{(2)}{{\cal E}} 
  := \stackrel{(2)}{\epsilon} 
  - 2 {\pounds}_{X} \stackrel{(1)}{\epsilon}
  - \left\{
    {\pounds}_{Y}
    -{\pounds}_{X}^{2}
  \right\} \epsilon
  , \nonumber
  \\
  \stackrel{(2)}{{\cal P}}
  &:=& \stackrel{(2)}{p}
  - 2 {\pounds}_{X} \stackrel{(1)}{p}
  - \left\{
    {\pounds}_{Y}
    -{\pounds}_{X}^{2}
  \right\} p
  , 
  \quad
  \label{eq:kouchan-016.18}
  \stackrel{(2)}{{\cal U}_{a}}
  := \stackrel{(2)}{(u_{a})}
  - 2 {\pounds}_{X} \stackrel{(1)}{u_{a}}
  - \left\{
    {\pounds}_{Y}
    -{\pounds}_{X}^{2}
  \right\} u_{a}.
  \nonumber
\end{eqnarray}
Through
$\bar{g}^{ab}\bar{u}_{a}\bar{u}_{b}$ $=$ $g^{ab}u_{a}u_{b}$ $=$
$-1$, the components of $\stackrel{(1)}{{\cal U}_{a}}$ and
$\stackrel{(2)}{{\cal U}_{a}}$ are given by
\begin{eqnarray}
  &&
  \stackrel{(1)}{{\cal U}_{a}}
  =
  - a \stackrel{(1)}{\Phi} (d\eta)_{a}
  + a \left( 
    D_{i} \stackrel{(1)}{v} + \stackrel{(1)}{{\cal V}_{i}}
  \right)(dx^{i})_{a},
  \quad
  \label{eq:kouchan-17.399}
  \stackrel{(2)}{{\cal U}_{a}} 
  =
  \stackrel{(2)}{{\cal U}_{\eta}} (d\eta)_{a}
  + a \left(
      D_{i} \stackrel{(2)}{v} 
    + \stackrel{(2)}{{\cal V}_{i}}
  \right) (dx^{i})_{a}
  , \\
  &&
  \stackrel{(2)}{{\cal U}_{\eta}}
  := 
  a \left\{
        \left(\stackrel{(1)}{\Phi}\right)^{2}
    -   \stackrel{(2)}{\Phi}
    - \left(
        D_{i}\stackrel{(1)}{v}
      + \stackrel{(1)}{{\cal V}_{i}}
      - \stackrel{(1)}{\nu_{i}}
    \right)
    \left(
        D^{i}\stackrel{(1)}{v}
      + \stackrel{(1)}{{\cal V}^{i}}
      - \stackrel{(1)}{\nu^{i}}
    \right)
  \right\}
\end{eqnarray}
where $D^{i}\stackrel{(1)}{{\cal V}_{i}}$ $=$
$D^{i}\stackrel{(2)}{{\cal V}_{i}}$ $=$ $0$.


We also expand the Einstein tensor as 
$\bar{G}_{a}^{\;\;b}$ $=$ $G_{a}^{\;\;b}$ $+$ 
$\lambda {}^{(1)}\!G_{a}^{\;\;b}$ $+$ 
$\frac{1}{2} \lambda^{2} {}^{(2)}\!G_{a}^{\;\;b}$.
From the decomposition of the first- and the second-order metric
perturbation into gauge-invariant parts and gauge-variant parts,
each order perturbation of the Einstein tensor is given by 
\begin{eqnarray}
  {}^{(1)}\!G_{a}^{\;\;b}
  =
  {}^{(1)}{\cal G}_{a}^{\;\;b}\left[{\cal H}\right]
  + {\pounds}_{X}G_{a}^{\;\;b}
  ,
  \quad
  {}^{(2)}\!G_{a}^{\;\;b}
  =
  {}^{(1)}{\cal G}_{a}^{\;\;b}\left[{\cal L}\right]
  + {}^{(2)}{\cal G}_{a}^{\;\;b}\left[{\cal H}, {\cal H}\right]
  + 2 {\pounds}_{X} {}^{(1)}\!G_{a}^{\;\;b}
  + \left\{
    {\pounds}_{Y}
    -{\pounds}_{X}^{2}
  \right\} G_{a}^{\;\;b}
\end{eqnarray}
as expected from Eqs.~(\ref{eq:matter-gauge-inv-def-1.0}).
Here, ${}^{(1)}{\cal G}_{a}^{\;\;b}\left[{\cal H}\right]$ and 
${}^{(1)}{\cal G}_{a}^{\;\;b}\left[{\cal L}\right]
+ {}^{(2)}{\cal G}_{a}^{\;\;b}\left[{\cal H}, {\cal H}\right]$
are gauge invariant parts of the first- and the second- order
perturbations of the Einstein tensor, respectively.
On the other hand, the energy momentum tensor of the perfect
fluid is also expanded as 
$\bar{T}_{a}^{\;\;b}$ $=$ $T_{a}^{\;\;b}$ $+$ 
$\lambda {}^{(1)}\!T_{a}^{\;\;b}$ $+$ 
$\frac{1}{2} \lambda^{2} {}^{(2)}\!T_{a}^{\;\;b}$
and ${}^{(1)}\!T_{a}^{\;\;b}$ and ${}^{(2)}\!T_{a}^{\;\;b}$ are
also given in the form
\begin{eqnarray}
  {}^{(1)}\!T_{a}^{\;\;b}
  =
  {}^{(1)}\!{\cal T}_{a}^{\;\;b}
  + {\pounds}_{X}T_{a}^{\;\;b}
  ,
  \quad
  {}^{(2)}\!T_{a}^{\;\;b}
  =
  {}^{(2)}\!{\cal T}_{a}^{\;\;b}
  + 2 {\pounds}_{X} {}^{(1)}\!T_{a}^{\;\;b}
  + \left\{
    {\pounds}_{Y}
    -{\pounds}_{X}^{2}
  \right\} T_{a}^{\;\;b}
\end{eqnarray}
through the definitions (\ref{eq:kouchan-016.18}) of the
gauge invariant variables of the fluid components.
Here, ${}^{(1)}\!{\cal T}_{a}^{\;\;b}$ and 
${}^{(2)}\!{\cal T}_{a}^{\;\;b}$ are gauge invariant part of the  
first- and the second-order perturbations of the energy momentum
tensor, respectively. 
Then, the first- and the second-order perturbations of the
Einstein equation are necessarily given in term of gauge
invariant variables:
\begin{eqnarray}
  \label{eq:linear-order-Einstein-equation}
  {}^{(1)}{\cal G}_{a}^{\;\;b}\left[{\cal H}\right]
  =
  8\pi G {}^{(1)}{\cal T}_{a}^{\;\;b},
  \quad
  {}^{(1)}{\cal G}_{a}^{\;\;b}\left[{\cal L}\right]
  + {}^{(2)}{\cal G}_{a}^{\;\;b}\left[{\cal H}, {\cal H}\right]
  =
  8\pi G \;\; {}^{(2)}{\cal T}_{a}^{\;\;b}. 
\end{eqnarray}


In the single perfect fluid case, the traceless scalar part of
the spatial component of the first equation in
Eq.(\ref{eq:linear-order-Einstein-equation}) yields
$\stackrel{(1)}{\Psi}$ $=$ $\stackrel{(1)}{\Phi}$ due to the
absence of the anisotropic stress in the first order
perturbation of the energy momentum tensor and the other
components of Eq.~(\ref{eq:linear-order-Einstein-equation}) give
well-known equations\cite{Bardeen-1980}.
We show the expression of the second-order perturbations of the
Einstein equation after imposing these first-order perturbations
of the Einstein equations.
Though we have derived all components of the second equation in
Eq.~(\ref{eq:linear-order-Einstein-equation}), we only show
their scalar parts of it for simplicity:
\begin{eqnarray}
  4\pi G a^{2} \stackrel{(2)}{{\cal E}}
  &=&
  \left(
    - 3 {\cal H} \partial_{\eta}
    +   \Delta
    + 3 K
    - 3 {\cal H}^{2}
  \right)
  \stackrel{(2)}{\Phi}
  - \Gamma_{0}
  +
  \frac{3}{2}
  \left(
    \Delta^{-1} D^{i}D_{j}\Gamma_{i}^{\;\;j}
    - \frac{1}{3} \Gamma_{k}^{\;\;k}
  \right)
  \nonumber\\
  && \quad
  -
  \frac{9}{2}
  {\cal H} \partial_{\eta}
  \left( \Delta + 3 K \right)^{-1}
  \left(
    \Delta^{-1} D^{i}D_{j}\Gamma_{i}^{\;\;j}
    - \frac{1}{3} \Gamma_{k}^{\;\;k}
  \right)
  \label{eq:kouchan-18.79}
  , \\
  8\pi G a^{2} (\epsilon + p) D_{i}\stackrel{(2)}{v} 
  &=&
  - 2 \partial_{\eta}D_{i}\stackrel{(2)}{\Phi}
  - 2 {\cal H} D_{i}\stackrel{(2)}{\Phi}
  + D_{i} \Delta^{-1} D^{k}\Gamma_{k}
  \nonumber\\
  && \quad
  - 3 \partial_{\eta}D_{i}
  \left( \Delta + 3 K \right)^{-1}
  \left(
    \Delta^{-1} D^{i}D_{j}\Gamma_{i}^{\;\;j}
    - \frac{1}{3} \Gamma_{k}^{\;\;k}
  \right)
  \label{eq:second-velocity-scalar-part-Einstein}
  , \\
  4 \pi G a^{2} \stackrel{(2)}{{\cal P}}
  &=&
  \left(
      \partial_{\eta}^{2} 
    + 3{\cal H} \partial_{\eta}
    - K
    + 2\partial_{\eta}{\cal H}
    + {\cal H}^{2}
  \right)
  \stackrel{(2)}{\Phi}
  -
  \frac{1}{2}
  \Delta^{-1} D^{i}D_{j}\Gamma_{i}^{\;\;j}
  \nonumber\\
  && \quad
  +
  \frac{3}{2}
  \left(
        \partial_{\eta}^{2} 
    + 2 {\cal H} \partial_{\eta}
  \right)
  \left( \Delta + 3 K \right)^{-1}
  \left(
    \Delta^{-1} D^{i}D_{j}\Gamma_{i}^{\;\;j}
    - \frac{1}{3} \Gamma_{k}^{\;\;k}
  \right)
  ,\\
  \label{eq:kouchan-18.80}
  \stackrel{(2)}{\Psi} - \stackrel{(2)}{\Phi}
  &=&
  \frac{3}{2}
  \left( \Delta + 3 K \right)^{-1}
  \left(
    \Delta^{-1} D^{i}D_{j}\Gamma_{i}^{\;\;j}
    - \frac{1}{3} \Gamma_{k}^{\;\;k}
  \right)
  .
  \label{eq:kouchan-18.65}
\end{eqnarray}
where ${\cal H}$ $:=$ $\partial_{\eta}a/a$.
$\Gamma_{0}$, $\Gamma_{i}$ and $\Gamma_{ij}$ in
Eqs.~(\ref{eq:kouchan-18.79})-(\ref{eq:kouchan-18.65}) are
defined by 
\begin{eqnarray}
  \Gamma_{0}
  &:=&
  +  8 \pi G a^{2} \left(\epsilon+p\right) D_{i}\stackrel{(1)}{v} D^{i}\stackrel{(1)}{v} 
  -  3 D_{k}\stackrel{(1)}{\Phi} D^{k}\stackrel{(1)}{\Phi}
  -  8 \stackrel{(1)}{\Phi} \Delta\stackrel{(1)}{\Phi}
  -  3 \left(\partial_{\eta}\stackrel{(1)}{\Phi}\right)^{2}
  - 12 \left(K + {\cal H}^{2}\right) \left(\stackrel{(1)}{\Phi}\right)^{2}
  \nonumber\\
  && 
  - 4 \left(
    \partial_{\eta}D_{i}\stackrel{(1)}{\Phi}+{\cal H} D_{i}\stackrel{(1)}{\Phi}
  \right) \stackrel{(1)}{{\cal V}^{i}} 
  - 2 {\cal H} D_{k}\stackrel{(1)}{\Phi} \stackrel{(1)}{\nu^{k}}
  + 8  \pi G a^{2} \left(\epsilon+p\right) \stackrel{(1)}{{\cal V}_{i}} \stackrel{(1)}{{\cal V}^{i}}
  + \frac{1}{2} D_{k}\stackrel{(1)}{\nu_{l}} D^{(k}\stackrel{(1)}{\nu^{l)}}
  + 3  {\cal H}^{2} \stackrel{(1)}{\nu^{k}} \stackrel{(1)}{\nu_{k}}
  \nonumber\\
  && 
  +             D_{l}D_{k}\stackrel{(1)}{\Phi} \stackrel{(1)}{\chi^{lk}}
  - 2 {\cal H} D^{k}\stackrel{(1)}{\nu^{l}} \stackrel{(1)}{\chi_{kl}}
  -\frac{1}{2}D^{k}\stackrel{(1)}{\nu^{l}}\partial_{\eta}\stackrel{(1)}{\chi_{lk}}
  \nonumber\\
  && 
  + \frac{1}{8} \partial_{\eta}\stackrel{(1)}{\chi_{lk}} \partial_{\eta}\stackrel{(1)}{\chi^{kl}}
  + {\cal H} \stackrel{(1)}{\chi_{kl}} \partial_{\eta}\stackrel{(1)}{\chi^{lk}}
  - \frac{1}{8} D_{k}\stackrel{(1)}{\chi_{lm}} D^{k}\stackrel{(1)}{\chi^{ml}}
  + \frac{1}{2} D_{k}\stackrel{(1)}{\chi_{lm}} D^{[l}\stackrel{(1)}{\chi^{k]m}}
  - \frac{1}{2} \stackrel{(1)}{\chi^{lm}} \left(\Delta-K\right)\stackrel{(1)}{\chi_{lm}}
  ,
  \nonumber\\
  \Gamma_{i}
  &:=&
  - 16 \pi G a^{2} \left(
    \stackrel{(1)}{{\cal E}} + \stackrel{(1)}{{\cal P}} 
  \right)
  D_{i}\stackrel{(1)}{v}
  +         12  {\cal H} \stackrel{(1)}{\Phi} D_{i}\stackrel{(1)}{\Phi}
  -          4  \stackrel{(1)}{\Phi} \partial_{\eta}D_{i}\stackrel{(1)}{\Phi}
  -          4  \partial_{\eta}\stackrel{(1)}{\Phi} D_{i}\stackrel{(1)}{\Phi}
  \nonumber\\
  &&
  - 16 \pi G a^{2} \left(
    \stackrel{(1)}{{\cal E}} + \stackrel{(1)}{{\cal P}} 
  \right)
  \stackrel{(1)}{{\cal V}_{i}}
  -          2  D^{j}\stackrel{(1)}{\Phi} D_{i}\stackrel{(1)}{\nu_{j}}
  +          2  D_{i}D^{j}\stackrel{(1)}{\Phi} \stackrel{(1)}{\nu_{j}}
  +          2  \Delta\stackrel{(1)}{\Phi} \stackrel{(1)}{\nu_{i}}
  +             \stackrel{(1)}{\Phi} \Delta\stackrel{(1)}{\nu_{i}}
  +          2  K \stackrel{(1)}{\Phi} \stackrel{(1)}{\nu_{i}}
  \nonumber\\
  &&
  -          4  {\cal H} \stackrel{(1)}{\nu^{j}} D_{i}\stackrel{(1)}{\nu_{j}}
  +          2  D^{j}\stackrel{(1)}{\Phi} \partial_{\eta}\stackrel{(1)}{\chi_{ji}}
  -          2  \partial_{\eta}D^{j}\stackrel{(1)}{\Psi} \stackrel{(1)}{\chi_{ij}}
  \nonumber\\
  &&
  +          2  D_{k}D_{[i}\stackrel{(1)}{\nu_{m]}} \stackrel{(1)}{\chi^{km}}
  +          2  D^{[k}\stackrel{(1)}{\nu^{j]}} D_{j}\stackrel{(1)}{\chi_{ki}}
  +          2  K \stackrel{(1)}{\nu^{j}} \stackrel{(1)}{\chi_{ij}}
  -             \stackrel{(1)}{\nu^{j}} \Delta\stackrel{(1)}{\chi_{ji}}
  - \frac{1}{2} \partial_{\eta}\stackrel{(1)}{\chi^{jk}} D_{i}\stackrel{(1)}{\chi_{kj}}
  +          2  \stackrel{(1)}{\chi^{kj}} \partial_{\eta}D_{[j}\stackrel{(1)}{\chi_{i]k}}
  ,
  \label{eq:kouchan-19.118}
  \\
  \Gamma_{ij}
  &:=&
    16 \pi G a^{2} \left( \epsilon + p \right) D_{i}\stackrel{(1)}{v} D_{j}\stackrel{(1)}{v}
  -  4 D_{i}\stackrel{(1)}{\Phi} D_{j}\stackrel{(1)}{\Phi}
  -  8 \stackrel{(1)}{\Phi} D_{i}D_{j}\stackrel{(1)}{\Phi}
  \nonumber\\
  && \quad\quad
  + \left\{
       6 D_{k}\stackrel{(1)}{\Phi} D^{k}\stackrel{(1)}{\Phi}
    +  8 \stackrel{(1)}{\Phi} \Delta\stackrel{(1)}{\Phi}
    +  2 \left(\partial_{\eta}\stackrel{(1)}{\Phi}\right)^{2}
    + 16 {\cal H} \stackrel{(1)}{\Phi} \partial_{\eta}\stackrel{(1)}{\Phi}
    + 8 \left(
      2 \partial_{\eta}{\cal H} + K + {\cal H}^{2}
    \right)
    \left(\stackrel{(1)}{\Phi}\right)^{2}
  \right\} \gamma_{ij}
  \nonumber\\
  && 
  + 32 \pi G a^{2} \left( \epsilon + p \right) D_{(i}\stackrel{(1)}{v} \stackrel{(1)}{{\cal V}_{j)}}
  -  4 \partial_{\eta}\stackrel{(1)}{\Phi} D_{(i}\stackrel{(1)}{\nu_{j)}}
  +  4 \partial_{\eta}D_{(i}\stackrel{(1)}{\Phi} \stackrel{(1)}{\nu_{j)}}
  + \left(
       4 \partial_{\eta}D_{k}\stackrel{(1)}{\Phi} \stackrel{(1)}{\nu^{k}}
    +  4 {\cal H} D_{k}\stackrel{(1)}{\Phi} \stackrel{(1)}{\nu^{k}}
  \right) \gamma_{ij}
  \nonumber\\
  && 
  + 16 \pi G a^{2} \left( \epsilon + p \right) \stackrel{(1)}{{\cal V}_{i}} \stackrel{(1)}{{\cal V}_{j}}
  -  2 \stackrel{(1)}{\nu^{k}} D_{k}D_{(i}\stackrel{(1)}{\nu_{j)}}
  +  2 \stackrel{(1)}{\nu_{k}} D_{i}D_{j}\stackrel{(1)}{\nu^{k}}
  +    D_{i}\stackrel{(1)}{\nu^{k}} D_{j}\stackrel{(1)}{\nu_{k}}
  +    D^{k}\stackrel{(1)}{\nu_{i}} D_{k}\stackrel{(1)}{\nu_{j}}
  \nonumber\\
  && \quad\quad
  + \left(
    -    D_{k}\stackrel{(1)}{\nu_{l}} D^{[k}\stackrel{(1)}{\nu^{l]}}
    -    D_{k}\stackrel{(1)}{\nu_{l}} D^{k}\stackrel{(1)}{\nu^{l}}
    -  2 \stackrel{(1)}{\nu_{k}} \Delta\stackrel{(1)}{\nu^{k}}
    -  4 \partial_{\eta}{\cal H} \stackrel{(1)}{\nu_{k}}\stackrel{(1)}{\nu^{k}}
    +  6 {\cal H}^{2} \stackrel{(1)}{\nu_{k}} \stackrel{(1)}{\nu^{k}}
  \right) \gamma_{ij}
  \nonumber\\
  && 
  -  4 {\cal H} \partial_{\eta}\stackrel{(1)}{\Phi} \stackrel{(1)}{\chi_{ij}}
  -  2 \partial_{\eta}^{2}\stackrel{(1)}{\Phi} \stackrel{(1)}{\chi_{ij}}
  -  4 D^{k}\stackrel{(1)}{\Phi} D_{(i}\stackrel{(1)}{\chi_{j)k}}
  +  4 D^{k}\stackrel{(1)}{\Phi} D_{k}\stackrel{(1)}{\chi_{ij}}
  -  8 K \stackrel{(1)}{\Phi} \stackrel{(1)}{\chi_{ij}}
  +  4 \stackrel{(1)}{\Phi} \Delta\stackrel{(1)}{\chi_{ij}}
  \nonumber\\
  && \quad\quad
  -  4 D^{k}D_{(i}\stackrel{(1)}{\Phi} \stackrel{(1)}{\chi_{j)k}}
  +  2 \Delta \stackrel{(1)}{\Phi} \stackrel{(1)}{\chi_{ij}}
  +  2 D_{l}D_{k}\stackrel{(1)}{\Phi} \stackrel{(1)}{\chi^{lk}} \gamma_{ij}
  \nonumber\\
  && 
  -  2 D^{k}\stackrel{(1)}{\nu_{(i}} \partial_{\eta}\stackrel{(1)}{\chi_{j)k}}
  -  2 \stackrel{(1)}{\nu^{k}} \partial_{\eta}D_{(i}\stackrel{(1)}{\chi_{j)k}}
  +  2 \stackrel{(1)}{\nu^{k}} \partial_{\eta}D_{k}\stackrel{(1)}{\chi_{ij}}
  +    D^{k}\stackrel{(1)}{\nu^{l}} \partial_{\eta}\stackrel{(1)}{\chi_{lk}} \gamma_{ij}
  \nonumber\\
  && 
  +             \partial_{\eta}\stackrel{(1)}{\chi_{ik}} \partial_{\eta}\stackrel{(1)}{\chi_{j}^{\;\;k}}
  +          2  D_{[l}\stackrel{(1)}{\chi_{k]i}} D^{k}\stackrel{(1)}{\chi_{j}^{\;\;l}}
  - \frac{1}{2} D_{j}\stackrel{(1)}{\chi_{lk}} D_{i}\stackrel{(1)}{\chi^{lk}}
  -             \stackrel{(1)}{\chi^{lm}} D_{i}D_{j}\stackrel{(1)}{\chi_{ml}}
  +          2  \stackrel{(1)}{\chi^{lm}} D_{l}D_{(i}\stackrel{(1)}{\chi_{j)m}}
  \nonumber\\
  && \quad\quad
  -             \stackrel{(1)}{\chi^{lm}} D_{m}D_{l}\stackrel{(1)}{\chi_{ij}}
  + \left(
    - \frac{3}{4} \partial_{\eta}\stackrel{(1)}{\chi_{lk}} \partial_{\eta}\stackrel{(1)}{\chi^{kl}}
    + \frac{3}{4} D_{k}\stackrel{(1)}{\chi_{lm}} D^{k}\stackrel{(1)}{\chi^{ml}}
    - \frac{1}{2} D_{k}\stackrel{(1)}{\chi_{lm}} D^{l}\stackrel{(1)}{\chi^{mk}}
    +             K \stackrel{(1)}{\chi_{lm}} \stackrel{(1)}{\chi^{lm}}
  \right) \gamma_{ij}
  \nonumber
  ,
\end{eqnarray}
and $\Gamma_{i}^{\;\;j}$ $:=$ $\gamma^{jk}\Gamma_{ik}$.
These equations
(\ref{eq:kouchan-18.79})-(\ref{eq:kouchan-18.65}) coincide with
the equations derived in Refs.\cite{KNs-cosmological} except for
the definition of the source terms $\Gamma_{0}$, $\Gamma_{i}$,
and $\Gamma_{ij}$.
Further, as shown in Refs.\cite{KNs-cosmological}, the equations
(\ref{eq:kouchan-18.79}) and (\ref{eq:kouchan-18.80}) are
reduced to the single equation for $\stackrel{(2)}{\Phi}$.
We also derived the similar equations in the case where the
matter content of the universe is a single scalar
field\cite{KNs-preparation}.


In summary, we have extended our formulation without ignoring
the first-order vector- and tensor-modes.
As the result, these equations imply that any types of
mode-coupling arise due to the second-order effects of the
Einstein equations, in principle. 
In some inflationary scenario, the tensor mode are also generated by
the quantum fluctuations.
This extension will be useful to clarify the evolution of the second
order perturbation in the existence of the first-order
tensor-mode.
Further, to apply this formulation to clarify the non-linear effects
in CMB physics\cite{Non-Gaussianity-in-CMB}, we have to extend our
formulation to multi-field system and to the Einstein-Boltzmann
system. 
These extensions will be one of our future works.



\begin{thebibliography}{99}
\bibitem{WMAP}
  C.L.~ Bennett et al., Astrophys. J. Suppl. Ser. {\bf 148},
  (2003), 1.
\bibitem{KNs-general}
  K.~Nakamura, Prog.~Theor.~Phys. {\bf 110} (2003), 723;
  K.~Nakamura, Prog.~Theor.~Phys. {\bf 113} (2005), 481.
\bibitem{KNs-cosmological}
  K.~Nakamura, Phys.~Rev.\ D {\bf 74} (2006), 101301(R);
  K.~Nakamura, Prog.~Theor.~Phys. {\bf 117} (2007), 17.
\bibitem{KNs-preparation}
  K.~Nakamura, in preparation.
\bibitem{Bardeen-1980}
  J.~M.~Bardeen, Phys. Rev. D {\bf 22} (1980), 1882;
  H.~Kodama and M.~Sasaki, Prog. Theor. Phys. Suppl. No.78, 1 (1984);
  V.~F.~Mukhanov, H.~A.~Feldman and R.~H.~Brandenberger,
  Phys. Rep. {\bf 215}, 203 (1992).
\bibitem{Non-Gaussianity-in-CMB}
  N.~Bartolo, S.~Matarrese and A.~Riotto, JCAP {\bf 0401}, 003 (2004);
  N.~Bartolo, S.~Matarrese and A.~Riotto, Phys. Rev. Lett.
  {\bf 93} (2004), 231301;
  N.~Bartolo, E.~Komatsu, S.~Matarrese and A.~Riotto,
  Phys. Rept. {\bf 402}, 103 (2004);
  N.~Bartolo, S.~Matarrese, and A.~Riotto, arXiv:astro-ph/0512481.
\end{thebibliography}
\end{document}